\documentstyle[12pt]{article}

\newcommand{\be}{\begin{equation}}
\newcommand{\ee}{\end{equation}}
\newcommand{\Dlt}{\Delta}
\newcommand{\dlt}{\delta}
\newcommand{\prt}{\partial}
\newcommand{\br}{{\bf r}}
\newcommand{\bk}{{\bf k}}

\newcommand{\bP}{{\bf P}}
\newcommand{\bp}{{\bf p}}
\newcommand{\bv}{{\bf v}}
\newcommand{\bt}{\beta}
\newcommand{\vp}{\varphi}
\newcommand{\ep}{\varepsilon}
\newcommand{\al}{\alpha}
\newcommand{\ra}{\rightarrow}
\newcommand{\sgm}{\sigma}

\newcommand{\gm}{\gamma}
\newcommand{\om}{\omega}
\newcommand{\Om}{\Omega}

\newcommand{\dgr}{\dagger}

\newcommand{\cF}{{\cal F}}

\begin{document}

\begin{center}
{\Large{\bf Bose-Einstein-condensed systems in random potentials} \\ [5mm]
V.I. Yukalov$^{1,2}$ and R. Graham$^1$} \\ [3mm]
{\it $^1$Fachbereich Physik, Universit\"at Duisburg-Essen, \\
47048 Duisburg, Germany \\ [2mm]

$^2$Bogolubov Laboratory of Theoretical Physics, \\
Joint Institute for Nuclear Research, Dubna 141980, Russia}

\end{center}

\vskip 2cm

\begin{abstract}

The properties of systems with Bose-Einstein condensate in external
time-independent random potentials are investigated in the frame of a
self-consistent stochastic mean-field approximation. General considerations
are presented, which are valid for finite temperatures, arbitrary strengths
of the interaction potential, and for arbitrarily strong disorder potentials.
The special case of a spatially uncorrelated random field is then treated
in more detail. It is shown that the system consists of three components,
condensed particles, uncondensed particles and a glassy density fraction,
but that the pure Bose glass phase with only a glassy density does not
appear. The theory predicts a first-order phase transition for increasing
disorder parameter, where the condensate fraction and the superfluid fraction
simultaneously jump to zero. The influence of disorder on the ground-state
energy, the stability conditions, the compressibility, the structure factor,
and the sound velocity are analyzed. The uniform ideal condensed gas is shown 
to be always stochastically unstable, in the sense that an infinitesimally 
weak disorder destroys the Bose-Einstein condensate, returning the system to 
the normal state. But the uniform Bose-condensed system with finite repulsive
interactions becomes stochastically stable and exists in a finite interval 
of the disorder parameter.

\end{abstract}

\vskip 1cm


\newpage

\section{Introduction}

The existence of the condensate fraction and its relation to the superfluid
fraction in random Bose media have been an intriguing research subject for
many years. First, the objects of interest have been $^4$He-filled porous
media, such as Vycor glasses, aerogel glasses, and grained powders [1,2].
Recently, the physics of dilute Bose gases has gained much interest (see
the books [3] and review articles [4--7]). Several experiments with
Bose-Einstein condensates in random potentials have been accomplished, and
different techniques of creating random fields have been proposed. For
example, random potentials can be formed by laser speckles [8,9] or by
randomly-varying magnetic fields in the close proximity of a current-carrying
wire [10]. Quasi-random potentials can also be created by using two-color
quasiperiodic noncommensurate optical lattices [11].

In the theory of disordered Bose systems, one considers two types of
models. Of one type are the lattice models characterized by a boson Hubbard
Hamiltonian with random site potentials. Such random potentials suppress
or may even can completely destroy the long-range order related to
Bose-Einstein condensates [12]. Fisher et al. [13] have suggested that
sufficiently strong disorder in a lattice leads to the appearance of a new
phase, different from insulating and superfluid phases. This is the Bose
glass phase, which is characterized by a finite compressibility, the absence
of a gap in the single particle spectrum, and a nonvanishing density of states
at zero energy. The phases in these lattice models can be classified [14] on
the basis of two order parameters, the condensate fraction $n_0$ and the
superfluid fraction $n_s$. In the {\it insulating phase}, $n_0=0$ and
$n_s=0$. For the {\it superfluid phase}, both order parameters are nonzero,
$n_0\neq 0$ and $n_s\neq 0$. And for the {\it Bose glass phase}, there is
$n_0\neq 0$, but there is no superfluidity, $n_s=0$. The occurrence of the
lattice Bose glass, arising between the insulating and superfluid phases,
has been investigated in several theoretical papers [15--21] and confirmed
in a recent experiment [11].

In a second class of models the disordered bosons can be thought of as being
immersed in  an initially uniform system in a random external potential,
with no regular lattices imposed. This type of  models was first studied
by Huang and Meng [22], who considered  the case of asymptotically weak
interactions and of asymptotically weak disorder in the Bogolubov
approximation. Their results were recovered by Giorgini et al.
[23] using the hydrodynamic approximation, which is mathematically equivalent
to the Bogolubov approximation. Lopatin and Vinokur [24] estimated the shift
of the critical temperature due to weak disorder in a weakly interacting gas,
which also was studied by Zobay [25], using renormalization group techniques.
If the results obtained for asymptotically weak disorder are formally
extended to strong disorder, then one comes [22,26] to the state, where
$n_0\neq 0$ but $n_s=0$, which corresponds to the Bose glass phase. However,
Monte Carlo simulations [27] for a gas with strong disorder, although it
confirmed that the superfluid fraction can be smaller than the condensate
fraction, found no presence of the Bose glass phase. Also, no Bose glass
was found in the random-phase approximation at zero temperature and
asymptotically weak interactions [28]. Instead, increasing disorder led
to a first-order transition from the superfluid to the normal phase. Thus,
the situation with Bose-condensed systems in random potentials is well
understood for the limit of weak interactions and weak disorder. However,
it remains controversial when the interactions and/or the disorder become
larger.

The aim of the present paper is to develop a new approach for treating
Bose-condensed systems in random potentials, when particle interactions
and strength of disorder can be arbitrary. We analyze the main properties
of the system and the influence of disorder and the interaction strength 
on these properties. In particular, the ideal uniform gas with 
Bose-Einstein condensate is shown to be stochastically unstable, in the 
sense that an infinitesimally weak random noise destroys the condensate, 
turning the system to the normal noncondensed state. The stochastic 
instability could be one of the reasons why the ideal Bose-Einstein 
condensation is not experimentally possible, and confining potentials 
and atomic interactions are necessary for the Bose-Einstein condensation 
to be realized in the laboratory. Nonvanishing repulsive atomic interactions 
stabilize the condensate, which can then exist in a finite domain of 
temperatures and of the disorder strength. At a temperature-dependent 
value of the latter the Bose-condensed system undergoes a first-order 
phase transition and transforms to the normal phase.

Throughout the paper a system of units is used, where $\hbar=1$ and $k_B=1$.

\section{System Hamiltonian}

The Hamiltonian energy operator is taken in the standard form
\be
\label{1}
\hat H[\psi] = \int \psi^\dgr(\br) \left [ -\; \frac{\nabla^2}{2m} +
\xi(\br)\right ]\psi(\br)\; d\br \; + \; \frac{1}{2}\; \Phi_0
\int \psi^\dgr(\br)\psi^\dgr(\br)\psi(\br)\psi(\br)\; d\br \; ,
\ee
in which $\psi(\br)=\psi(\br,t)$ is the Bose field operator, $\xi(\br)$ is
a random external potential, and the particle interaction strength
\be
\label{2}
\Phi_0 = 4\pi\; \frac{a_s}{m}
\ee
is expressed through the scattering length $a_s$ and particle mass $m$.

The averaging over the random potentials will be denoted by the double angle
brackets $\ll\ldots\gg$. The distribution over the random fields is assumed
to be zero-centered, so that
\be
\label{3}
\ll \xi(\br) \gg \; = \; 0 \; .
\ee
The stochastic average
\be
\label{4}
\ll \xi(\br)\xi(\br') \gg \; = \; R(\br-\br')
\ee
defines the correlation function $R(\br)$. The random potential and the
correlation function are supposed to be real and the latter is also symmetric,
such that
\be
\label{5}
\xi^*(\br) = \xi(\br)\; , \qquad R^*(\br)=R(-\br) = R(\br) \; .
\ee
Therefore their Fourier transforms enjoy the properties
\be
\label{6}
\xi_k^* = \xi_{-k} \; , \qquad R^*_k = R_{-k} = R_k \; .
\ee

The Fourier transform $\xi_k$ possesses also the important property 
$\xi_k\ra 0$, when $k\ra\infty$ as explained in the Appendix
A. In the Fourier representation, Eq. (4) reduces to
\be
\label{7}
\ll \xi^*_k\xi_p\gg \; = \; \dlt_{kp} R_k V \; .
\ee
For the particular case of white noise, when
\be
\label{8}
R(\br) = R_0\dlt(\br) \; ,
\ee
one has
\be
\label{9}
\ll \xi^*_k \xi_p \gg  \; = \; \dlt_{kp} R_0 V \; .
\ee

The main part of the present paper will not depend on the particular
type of the distribution over the random potentials, and hence on the
concrete choice of the correlation functions (4) and (7). But at the
final stage, in order to illustrate practical calculations, we shall
specialize to the white noise characterized by Eqs. (8) and (9).

All operators from the algebra of local observables are functionals of
the field operators $\psi(\br)$ and $\psi^\dgr(\br)$ and of the random
variable $\xi(\br)$. This implies that there are two kinds of averages.
One kind is the stochastic average $\ll\hat A\gg$ over the distribution
of the random potentials. And another one is the quantum average with
respect to a Hamiltonian $H$, which is denoted as
\be
\label{10}
<\hat A>_H \; \equiv\; {\rm Tr}\;\hat\rho \hat A \; ,
\ee
with the statistical operator
\be
\label{11}
\hat\rho  = \frac{\exp(-\bt H)}{{\rm Tr}\exp(-\bt H)} \; .
\ee
Here the Hamiltonian $H$ includes, but is, in general, different from
$\hat H$ and remains to be specified below. $\bt\equiv 1/T$ is the
inverse temperature and the trace is over the
Fock space $\cF(\psi)$ generated by the related field operators [29,30].
The total average will be denoted as
\be
\label{12}
<\hat A> \; \equiv \; \ll {\rm Tr}\; \hat\rho \hat A\gg \; .
\ee

To describe a Bose-condensed system, where the global gauge symmetry is
broken, one employs the Bogolubov shift
\be
\label{13}
\psi(\br)\; \longrightarrow \; \hat\psi(\br) \equiv
\eta(\br) + \psi_1(\br) \; ,
\ee
where $\eta(\br)$ is the condensate wave function.
 The field variable $\eta(\br)$ and the operator $\psi_1(\br)$ are
taken as linearly independent and orthogonal to each other,
\be
\label{14}
\int \eta^*(\br)\psi_1(\br) \; d\br = 0 \; .
\ee
$\psi_1(\br)$ is
the operator of uncondensed particles, satisfying the Bose commutation
relations [31--33]. The condensate function is normalized to a fixed, still
undetermined, positive value $N_0$, the number of condensed particles
\be
\label{15}
N_0 = \int |\eta(\br)|^2 \; d\br \; .
\ee
The physical value of $N_0$ must then be chosen by minimizing the
thermodynamic potential. The number of uncondensed particles $N_1=N-N_0$
is given by the average
\be
\label{16}
N_1 \; = \; <\hat N_1>
\ee
of the number-of-particle operator
\be
\label{17}
\hat N_1 \equiv \int \psi_1^\dgr(\br) \psi_1(\br) \; d\br \; .
\ee
The total number of particles in the system is
\be
\label{18}
N \; = \; <\hat N>\; = \; N_0 + N_1 \; ,
\ee
with the operator
\be
\label{19}
\hat N \equiv \int \hat\psi^\dgr(\br) \hat\psi(\br)\; d\br =
N_0 + \hat N_1 \; ,
\ee
in which $\hat\psi(\br)$ is the shifted field operator (13).

According to these definitions, for the correct description of a Bose-condensed system,
which would be
self-consistent in any approximation, one therefore has to employ a representative
ensemble [34] taking into account the normalization conditions (15) {\it and} (16) or (18).
This requires [34--36] to use the grand Hamiltonian
\be
\label{20}
H \equiv \hat H - \mu_0 N_0 - \mu_1 \hat N_1 \; ,
\ee
where $\hat H = \hat H[\hat\psi]$, while $\mu_0$ and $\mu_1$ are the
Lagrange multipliers guaranteeing the validity of normalizations (15)
and (16). Here we shall consider an equilibrium system, but a similar
representative ensemble can also be defined for nonequilibrium
Bose-condensed systems [34,37].

\section{Thermodynamic Potential}

For the frozen disorder, the grand thermodynamic potential is
\be
\label{21}
\Om = - T \ll \ln{\rm Tr} e^{-\bt H} \gg \; .
\ee
To provide thermodynamic stability, potential (21) is to be minimal with
respect to the number of condensed particles,
\be
\label{22}
\frac{\prt\Om}{\prt N_0} = 0 \; , \qquad
\frac{\prt^2\Om}{\prt N_0^2} > 0 \; .
\ee
The system free energy can be defined as
\be
\label{23}
F = \Om + \mu_0 N_0 + \mu_1 N_1 \; .
\ee
At the same time, keeping in mind that in standard experiments only the
total number of particles $N$ is fixed, but not $N_0$ and $N_1$ separately,
we may write
\be
\label{24}
F = \Om + \mu N \; .
\ee
Comparing Eqs. (23) and (24) yields the definition of the system chemical
potential
\be
\label{25}
\mu \equiv \mu_0 n_0 + \mu_1 n_1 \; ,
\ee
in which $n_0\equiv N_0/N$, $n_1\equiv N_1/N$ are the corresponding
fractions of particles, satisfying the normalization condition $n_0+n_1=1$.

It is worth noting that, instead of working with the grand ensemble
containing two Lagrange multipliers, we could resort to the canonical
ensemble with no Lagrange multipliers but with two constraints that
are to be satisfied at each step of any calculational procedure. One
constraint is that the number of condensed particles $N_0=N_0(T,N)$ be
fixed by stability conditions, while the total number of particles $N$
be kept fixed at each step, but not solely on average. Such a canonical
ensemble could, probably, be realized with the help of the Girardeau-Arnowitt
representation [38]. However, a weak point of the latter is not only that
it leads to rather cumbersome calculations but, most importantly, that it
does not allow simple self-consistent approximations. For instance, it is
well known that the Hartree-Fock-Bogolubov (HFB) approximation is not
self-consistent in the frame of the Girardeau-Arnowitt representation,
yielding  an unphysical gap in the spectrum [38] for a uniform Bose system.
Girardeau [39] stressed the necessity to deal with the complete Hamiltonian
in order to make the canonical-ensemble approach self-consistent and to
remove the unphysical gap. Indeed, Takano showed [40] that this could really
be done at least in principle, if one would use all terms of the Hamiltonian.
However this necessity  makes the problem practically unsolvable: in general,
an exact solution for the problem is not known, and as soon as an approximation
is involved, one confronts the danger of getting not self-consistent results
[41]. Contrary to this, relaxing the imposed constraints, by introducing the
corresponding Lagrange multipliers, being mathematically equivalent, makes
all calculations much simpler, at the same time preserving the theory
self-consistency for any given approximation [34--37].

In order to calculate the thermodynamic potential (21) for the frozen
disorder, one often takes recource to the so-called replica trick, as is
used in the theory of spin glasses [42]. Here we shall employ another approach,
based on the method of separation of variables. The idea of this method is
as follows. The main aim is to transform the given Hamiltonian $H$ to a
separable form
\be
\label{26}
H_{sep} = H_q + H_\xi \; ,
\ee
in which $H_q$ depends only on quantum variables, while $H_\xi$ depends
only on classical stochastic variables. Such a transformation can be achieved by means
of canonical transformations and some simplifications. Then the corresponding
thermodynamic potential
\be
\label{27}
\Om_{sep} \equiv -T \ll \ln{\rm Tr} e^{-\bt H_{sep}} \gg
\ee
reduces to the sum
\be
\label{28}
\Om_{sep} = - T\ln{\rm Tr} e^{-\bt H_q} + \ll H_\xi \gg \; ,
\ee
in which the manipulations with quantum and stochastic variables are
separated. If the separable Hamiltonian (26) does not exactly represent the
initial $H$, so that
\be
\label{29}
H = H_{sep} + \hat h \qquad (\hat h \equiv H - H_{sep} ) \; ,
\ee
then corrections to the thermodynamic potential can be obtained by
perturbation theory with respect to $\hat h$, giving in the second order
\be
\label{30}
\Om = \Om_{sep}\; + <\hat h>  -\bt\Dlt^2(\hat h) \; ,
\ee
where $\Dlt^2(\hat h)$ is the dispersion
$\Dlt^2(\hat h)\equiv<\hat h^2>-< \hat h>^2$. In agreement with Eq. (30),
one has $\Om\leq\Om_{sep}+<\hat h>$, which is the Gibbs-Bogolubov inequality.

The method of separation of variables has no need for the replica trick. The
derivation of the separable Hamiltonian (26) can be accomplished by means of
decouplings and canonical transformations and does not require the existence
of small parameters. All essential nonlinearities with respect to particle
interactions and disorder strength can be preserved in the Hamiltonian (26).
The use of the Gibbs-Bogolubov inequality, mentioned above, can be done in
the standard variational way, by minimizing the right-hand side of this
inequality, which again does not require the existence of small parameters.
Therefore this method makes it possible to consider strong interactions and
strong disorder.

\section{Stochastic Quantization}

According to Eq. (3), the external random potential is zero on average. This
allows us to treat the condensate wave function, which is the system order
parameter, as uniform, so that [22]
\be
\label{31}
\eta(\br) \; = \; <\hat\psi(\br)>\; = \; \sqrt{\rho_0} \; ,
\ee
where $\hat\psi(\br)$ is the shifted field operator, $\rho_0\equiv N_0/V$
is the condensate density, and the total average (12) is assumed. In
agreement with Eq. (13), one has
\be
\label{32}
<\psi_1(\br)> \; = \; 0 \; .
\ee

Expanding the field operators of uncondensed particles in plane waves,
we represent the grand Hamiltonian (20) as the sum
\be
\label{33}
H = \sum_{n=0}^4 H^{(n)} + H_{ext} \; .
\ee
Here the zero-order term
\be
\label{34}
H^{(0)} = \left ( \frac{1}{2}\; \rho_0 \Phi_0 -
\mu_0 \right ) N_0
\ee
does not contain the operators of uncondensed particles. For the first-order
term, because of the property (14), we get
\be
\label{35}
H^{(1)} = 0 \; .
\ee
The term of second order, with respect to the operators $a_k$, becomes
\be
\label{36}
H^{(2)} = \sum_{k\neq 0} \left ( \frac{k^2}{2m}  + 2\rho_0 \Phi_0 -
\mu_1 \right ) a_k^\dgr a_k \;
+ \; \frac{1}{2} \; \sum_{k\neq 0} \rho_0 \Phi_0 \left (
a_k^\dgr a_{-k}^\dgr + a_{-k} a_k \right ) \; .
\ee
For the third-order term, we have
\be
\label{37}
H^{(3)} = \sqrt{\frac{\rho_0}{V}} \; {\sum_{k,p}}' \; \Phi_0 \left (
a_k^\dgr a_{k+p} a_{-p} + a_{-p}^\dgr a_{k+p}^\dgr a_k \right ) \; ,
\ee
where the prime on the summation symbol implies that $\bk\neq 0$, $\bp\neq 0$,
$\bk+\bp\neq 0$. The fourth-order term is
\be
\label{38}
H^{(4)} = \frac{1}{2V} \; \sum_q \; {\sum_{k,p}}' \;
\Phi_0 a_k^\dgr a_p^\dgr a_{k-q} a_{p+q} \; ,
\ee
where the prime on the summation sign means that $\bk\neq 0$, $\bp\neq 0$,
$\bk-{\bf q}\neq 0$, $\bp+{\bf q}\neq 0$. The last term in Eq. (33) corresponds
to the action of the external random field, given by the expression
\be
\label{39}
H_{ext} = \rho_0 \xi_0 + \sqrt{\frac{\rho_0}{V}} \;
\sum_{k\neq 0} \left ( a_k^\dgr \xi_k + \xi_k^* a_k \right ) +
\frac{1}{V} \; \sum_{k,p(\neq 0)} a_k^\dgr a_p \xi_{k-p} \; .
\ee

When one assumes asymptotically weak interactions, one omits the terms
$H^{(3)}$ and $H^{(4)}$, thus, coming to the Bogolubov approximation [31--33].
Since we aim at considering arbitrarily strong interactions, we have to keep
all terms of Hamiltonian (33). But we may simplify the terms $H^{(3)}$ and
$H^{(4)}$ by means of the Hartree-Fock-Bogolubov (HFB) approximation [35,36].
Then, we get
\be
\label{40}
H^{(3)} = 0 \; .
\ee
To express the result for the term $H^{(4)}$ in a compact form, we introduce
the normal average
\be
\label{41}
n_k \; \equiv \; <a_k^\dgr a_k > \; ,
\ee
which is the momentum distribution of atoms, and the anomalous average
\be
\label{42}
\sgm_k \; \equiv \; <a_k a_{-k} > \; .
\ee
The quantity $|\sgm_k|$ can be interpreted as the momentum distribution of
paired particles [35]. Then the density of uncondensed particles is
\be
\label{43}
\rho_1 = \frac{1}{V} \; \sum_{k\neq 0} n_k \; ,
\ee
while the sum
\be
\label{44}
\sgm_1 = \frac{1}{V} \; \sum_{k\neq 0} \sgm_k
\ee
gives the density $|\sgm_1|$ of paired particles. Applying the mean-field approximation
we find from Eq. (38)
$$
H^{(4)} = \sum_{k\neq 0} \rho_1 \Phi_0 \left ( a_k^\dgr a_k \; - \;
\frac{1}{2}\; n_k \right ) +
$$
\be
\label{45}
+ \frac{1}{V} \sum_{k,p(\neq 0)} \Phi_0 \left [
n_{k+p} a_p^\dgr a_p + \frac{1}{2}\left (
\sgm_{k+p}a_p^\dgr a_{-p}^\dgr + \sgm_{k+p}^* a_{-p} a_p \right ) -
\; \frac{1}{2} \left ( n_{k+p} n_p + \sgm_{k+p}\sgm_p^* \right )
\right ] \; .
\ee

A special care has to be taken in reorganizing expression (39) describing
the interaction of atoms with external random fields. The second term in
Eq. (39) corresponds to {\it linear} interactions between random fields and
atoms, while the third term describes {\it nonlinear} interactions. If one
omits the third term, as has been done by Huang and Meng [22], thus, keeping
solely the linear interactions, then one limits oneself by weak disorder.
Since our aim is to consider arbitrarily strong disorder, we need to keep this
term. The difficulty with treating the nonlinear term in Eq. (39) is that, in
the mean-field approximation, it is zero on the average, as far as
\be
\label{46}
<a_k^\dgr a_p \xi_{k-p}> \; \approx \;
< a_k^\dgr a_p><\xi_{k-p}> \; = \; 0 \; .
\ee
If we would treat this term in the simple mean-field manner replacing
$a_k^\dgr a_p\xi_{k-p}$ by
$$
<a_k^\dgr a_p>\xi_{k-p} + a_k^\dgr a_p <\xi_{k-p}> -
<a_k^\dgr a_p><\xi_{k-p}> \; = \; \dlt_{kp} n_k \xi_0 \; ,
$$
we would kill all quantum effects, reducing the term to the trivial form.
The way out of this problem is to employ a more refined approximation.

We shall use the ideas of the {\it stochastic mean-field approximation},
which has been applied to accurately treat quantum and stochastic effects in
systems interacting with electromagnetic fields [43] and in spin systems
[44,45]. In considering these systems, one encounters the same type of the
difficulty. If one uses the simple mean-field approximation, often called
semiclassical, then quantum and random effects are washed out, which may
lead to principally wrong results. To accurately take account of the
latter effects, the mean-field approximation is to be modified [43--45].

Let us remember that we have two types of averages for any operator
$\hat A$. The stochastic average $\ll\hat A\gg$ and the quantum average
$<\hat A>_H$ defined in Eq. (10). The operators of uncondensed particles
$a_k$ and $a_k^\dgr$ are, strictly speaking, functions of the random fields
$\xi_k$. We may separate the quantum and stochastic averages and consider
the quantum average
\be
\label{47}
\al_k \; \equiv \; < a_k>_H \; ,
\ee
which is a function of the random fields. This quantity $\al_k$ is not zero,
even though its total average
\be
\label{48}
<a_k>\; = \; \ll \al_k \gg \; = \; 0
\ee
is of course zero, according to Eq. (32). In the nonlinear term of Eq.
(39), in the spirit of the stochastic mean-field approximation [43--45],
we now make a mean-field type decoupling with respect to the quantum averaging
only, not with respect to the stochastic average, that is, we write
\be
\label{49}
a_k^\dgr a_p \xi_{k-p} =\left ( a_k^\dgr <a_p>_H +
<a_k^\dgr>_H a_p - <a_k^\dgr>_H <a_p>_H \right ) \xi_{k-p} \; .
\ee
One may notice that if we would employ in decoupling (49) the total
averages of type (12), instead of the quantum averages of type (10), then
the left-hand side of Eq. (49), according to Eq. (48), would be reduced
to zero, similar to Eq. (46). In order to retain the influence of the
left-hand side term of Eq. (49), we invoke here not the total but only
the {\it quantum averages}. Using the latter, instead of the total averages
(12), makes decoupling (49) more general, thus, allowing us to retain the
influence of nonlinear stochastic terms [43--45]. Let us also define the
stochastic field
\be
\label{50}
\vp_k \equiv \frac{\sqrt{N_0}}{V}\; \xi_k +
\frac{1}{V} \sum_{p\neq 0} \al_p \xi_{k-p} \; .
\ee
Then the random-field Hamiltonian (39) transforms to
\be
\label{51}
H_{ext} = \rho_0 \xi_0 + \sum_{k\neq 0} \left (
a_k^\dgr \vp_k + \vp_k^* a_k \right ) \; - \;
\frac{1}{V} \sum_{k,p(\neq 0)} \al_k^* \al_p \xi_{k-p} \; ,
\ee
where $\xi_0=\int\xi(\br)d\br$.

Finally, introducing the notation
\be
\label{52}
\om_k \equiv \frac{k^2}{2m} + 2\rho \Phi_0 - \mu_1 \; ,
\ee
where $\rho\equiv\rho_0+\rho_1$ is the total particle density, and defining
\be
\label{53}
\Dlt \equiv (\rho_0+\sgm_1) \Phi_0 \; ,
\ee
we obtain for Hamiltonian (33) the form
\be
\label{54}
H =E_{HFB} + \sum_{k\neq 0} \om_k a_k^\dgr a_k \;
+ \; \frac{1}{2} \sum_{k\neq 0} \Dlt\left ( a_k^\dgr a_{-k}^\dgr +
a_{-k} a_k \right ) + H_{ext} \; ,
\ee
in which
\be
\label{55}
E_{HFB} \equiv H^{(0)} \; - \; \frac{1}{2\rho} \left (
2\rho_1^2 + \sgm_1^2 \right ) \Phi_0 N
\ee
and $H_{ext}$ is given by Eq. (51). It is worth emphasizing that Hamiltonian
(54) has the mean-field form with respect to the field operators $a_k$, but
it contains, via $H_{ext}$, the nonlinear terms with respect to the random variables $\xi_k$,
$\al_k$, and $\vp_k$. The latter allows us to consider disorder of arbitrary
strength.

\section{Separation of Variables}

Quantum and stochastic variables in the Hamiltonian (54) are yet intermixed.
To separate them, we shall use the method of canonical transformations. First,
we employ the usual Bogolubov canonical transformation
\be
\label{56}
a_k = u_k b_k + v_{-k}^* b_{-k}^\dgr \; , \qquad
a_k^\dgr = u_k^* b_k^\dgr + v_{-k} b_{-k} \; .
\ee
Using these in Eq. (54), we get
\be
\label{57}
H = E_B + \sum_{k\neq 0} \ep_k b_k^\dgr b_k + H_{ext} \; ,
\ee
where
\be
\label{58}
E_B \equiv E_{HFB}  + \frac{1}{2} \sum_{k\neq 0} (\ep_k - \om_k )
\ee
and $\ep_k$ is the Bogolubov spectrum
\be
\label{59}
\ep_k =\sqrt{\om_k^2 - \Dlt^2} \; .
\ee
Equation (51), containing random fields, now becomes
\be
\label{60}
H_{ext} =\rho_0\xi_0 + \sum_{k\neq 0} \left ( b_k^\dgr D_k +
D_k^* b_k \right ) \; - \;
\frac{1}{V} \; \sum_{k,p(\neq 0)} \al_k^* \al_p \xi_{k-p} \; ,
\ee
where
\be
\label{61}
D_k \equiv \left ( u_k^* +  v_k^* \right ) \vp_k \; .
\ee
The coefficient functions in transformation (56) are defined by the equations
\be
\label{62}
u_k^2 = \frac{\sqrt{\ep_k^2+\Dlt^2}+\ep_k}{2\ep_k} =
\frac{\om_k+\ep_k}{2\ep_k} \; , \qquad
v_k^2 = \frac{\sqrt{\ep_k^2+\Dlt^2}-\ep_k}{2\ep_k} =
\frac{\om_k-\ep_k}{2\ep_k} \; .
\ee

Then we apply another canonical transformation
\be
\label{63}
b_k = \hat b_k \; - \; \frac{D_k}{\ep_k} \; , \qquad
b_k^\dgr = \hat b_k^\dgr \; - \; \frac{D_k^*}{\ep_k} \; ,
\ee
which transforms Hamiltonian (60) into
\be
\label{64}
H = E_B + \sum_{k\neq 0} \ep_k \hat b_k^\dgr \hat b_k + H_\xi \; ,
\ee
where $E_B$ is the nonoperator part (58), the second term does not depend
on stochastic variables, while the last term
\be
\label{65}
H_\xi = \rho_0 \xi_0  -
\sum_{k\neq 0} \; \frac{\vp_k^*\vp_k}{\om_k+\Dlt} \; - \;
\frac{1}{V} \sum_{k,p(\neq 0)} \al_k^* \al_p \xi_{k-p}
\ee
contains only stochastic fields, but no quantum variables.

In that way, the quantum operator variables $\hat b_k$ and
$\hat b_k^\dgr$ and the stochastic fields $\xi_k$, $\al_k$, and
$\vp_k$ are separated in Hamiltonian (64). This will allow us to
calculate different averages and to analyze the influence of random
fields on the system.

\section{Random Fields}

Let us, first, consider the Bogolubov spectrum (59). As is seen, it
does not explicitly depend on the random fields, thus, representing
the spectrum of collective excitations for a system that is uniform
on the average. For a uniform system, there exists the Hugenholtz-Pines
theorem [46,47] requiring that the spectrum be gapless, so that
\be
\label{66}
\lim_{k\ra 0} \ep_k = 0 \; , \qquad \ep_k \geq 0 \; .
\ee
Then, from Eqs. (52), (53), and (59), it follows that
\be
\label{67}
\mu_1 = (\rho + \rho_1 - \sgm_1)\Phi_0 \; .
\ee
As a result, Eq. (52) reduces to
\be
\label{68}
\om_k = \frac{k^2}{2m} + \Dlt \; .
\ee
The Bogolubov spectrum (59) acquires the form
\be
\label{69}
\ep_k = \sqrt{(ck)^2 + \left ( \frac{k^2}{2m}\right )^2} \; ,
\ee
in which the sound velocity
\be
\label{70}
c \equiv \sqrt{\frac{\Dlt}{m} }
\ee
is expressed through the quantity
\be
\label{71}
\Dlt \equiv mc^2 = (\rho_0 +\sgm_1) \Phi_0 \; ,
\ee
following from Eq. (53).

Another way of deriving Eq. (67) and, respectively, the Bogolubov spectrum
(69) is as follows. We may consider the equations of motion for the matrix
Green function $G(\bk,\om)=[G_{\al\bt}(\bk,\om)]$ as has been done by
Bogolubov [48]. The presence of the random-field Hamiltonian (39) contributes
to these equations with the terms all of which, in the mean-field approximation,
can be set zero, in agreement with Eq. (46). For the Green functions, one has
the Bogolubov theorem [48]
$$
| G_{11}(\bk,0) -  G_{12}(\bk,0)| \geq \frac{mn_0}{k^2} \; ,
$$
from which the Hugenholtz-Pines relation
$\mu_1=\Sigma_{11}(0,0)-\Sigma_{12}(0,0)$
follows, where $\Sigma(\bk,\om)=[\Sigma_{\al\bt}(\bk,\om)]$ is the matrix
self-energy. In the HFB approximation, we have $\Sigma_{11}(0,0)=2\rho\Phi_0$
and $\Sigma_{12}(0,0)=(\rho_0+\sgm_1)\Phi_0$. This gives us exactly the same
equation (67).

Combining the canonical transformations (56) and (63), we get
\be
\label{72}
a_k = u_k\hat b_k + v_{-k}^*\hat b_{-k}^\dgr -
\frac{|u_k+v_k|^2}{\ep_k}\; \vp_k \; , \qquad
a_k^\dgr = u_k^*\hat b_k^\dgr + v_{-k} \hat b_{-k} -
\frac{|u_k+v_k|^2}{\ep_k}\; \vp_k^* \; .
\ee
Because of the form of the Hamiltonian (64), one has
$<\hat b_k>=<\hat b_k \hat b_p>=0$. Then, from Eqs. (47) and (72), we find
\be
\label{73}
\al_k = -\; \frac{\vp_k}{\om_k+\Dlt} \; .
\ee
Hence,
\be
\label{74}
\ll |\al_k|^2\gg \; = \; \frac{\ll|\vp_k|^2\gg}{(\om_k+\Dlt)^2} \; .
\ee
By Eq. (48), we also have $<\vp_k>=\ll \vp_k\gg = 0$. Substituting
relation (73) into Eq. (50), we come to the equation
\be
\label{75}
\vp_k = \frac{\sqrt{N_0}}{V}\;\xi_k \; - \;
\frac{1}{V} \sum_{p\neq 0} \frac{\xi_{k-p}\vp_p}{\om_p+\Dlt}
\ee
defining the random field $\vp_k$. This is a Fredholm equation of
the second kind.

Using Hamiltonian (64), it is straightforward to get the momentum
distribution of quasiparticles
\be
\label{76}
\pi_k \; \equiv \; <\hat b_k^\dgr \hat b_k>\; = \left ( e^{\bt\ep_k} -1
\right )^{-1} = \frac{1}{2}{\rm coth}\left ( \frac{\ep_k}{2T}\right )
- \; \frac{1}{2} \; .
\ee
For the momentum distribution of atoms (41), we find
\be
\label{77}
n_k =\left ( u_k^2 + v_k^2\right ) \pi_k + v_k^2 +
\ll |\al_k|^2 \gg
\ee
and for the anomalous average (42), we have
\be
\label{78}
\sgm_k = (1+2\pi_k) u_k v_k + \ll |\al_k|^2\gg \; .
\ee
With Eqs. (62) and (76), we finally obtain the normal average
\be
\label{79}
n_k = \frac{\om_k}{2\ep_k}\; {\rm coth}\left ( \frac{\ep_k}{2T}
\right ) - \; \frac{1}{2} \;  + \ll |\al_k|^2 \gg
\ee
and the anomalous average
\be
\label{80}
\sgm_k = -\; \frac{\Dlt}{2\ep_k}\; {\rm coth}\left (
\frac{\ep_k}{2T}\right ) + \ll |\al_k|^2 \gg \; .
\ee
The contribution of the random potential comes through the last terms
in Eqs. (79) and (80). These terms are related to the random field $\vp_k$
by means of Eqs. (73) and (74). And the random field $\vp_k$ is defined as
the solution of the Fredholm equation (75).

\section{Glassy Fraction}

In order to elucidate the physical meaning of the terms, induced by the
random potential, let us draw some analogies with the theory of spin
glasses [42]. For the Bose system, we may define an order parameter, which
is the analogue of the Edwards-Anderson order parameter in spin glasses
[42]. To this end, we recall that the total average $<\psi_1>=0$, according
to Eq. (32). But, separating the quantum and stochastic averages, we can
introduce the {\it density of the glassy fraction}
\be
\label{81}
\rho_G \equiv \frac{1}{V} \int \ll |<\psi_1(\br)>_H|^2 \gg d\br \; .
\ee
Passing to the Fourier transform of $\psi_1(\br)$ and using Eqs. (72),
we reduce Eq. (81) to
\be
\label{82}
\rho_G = \frac{1}{V} \sum_{k\neq 0} \ll |\al_k|^2 \gg \; .
\ee
Consequently, the meaning of the quantity
\be
\label{83}
n_G(\bk) \; \equiv \; \ll |\al_k|^2 \gg \; = \;
\frac{\ll|\vp_k|^2\gg}{(\om_k+mc^2)^2}
\ee
is the momentum distribution of the glassy fraction.
We may assume that the nominator of Eq. (83) is not increasing with $k$.
However, its denominator, according to Eq. (68), increases with
$k$ as $k^4$. Hence, distribution (83) is a rapidly decreasing function of
$k$, with its maximum at $k=0$, where
\be
\label{84}
n_G(0)  = \frac{\ll|\vp_0|^2\gg}{4(mc^2)^2} \; .
\ee

The glassy density (82), using relation (74), can be represented as
\be
\label{85}
\rho_G = \int \frac{\ll|\vp_k|^2\gg}{(\om_k+\Dlt)^2}\;
\frac{d\bk}{(2\pi)^3} \; .
\ee
Since the integrand in Eq. (85) falls off rapidly and $\ll|\vp_k|^2\gg$
is slowly varying with $k$, we may substitute
$\ll|\vp_0|^2\gg$ instead of $\ll|\vp_k|^2\gg$, which gives
\be
\label{86}
\rho_G = \frac{(mc)^3}{\pi}\; n_G(0) \; .
\ee
For the dimensionless {\it glassy fraction}, we then have
\be
\label{87}
n_G \equiv \frac{\rho_G}{\rho} = \frac{(mc)^3}{\pi\rho}\; n_G(0) \; .
\ee

Let us consider the glassy density matrix
\be
\label{88}
\rho_G(\br_1,\br_2) \equiv \int n_G(\bk) e^{i\bk\cdot\br_{12}} \;
\frac{d\bk}{(2\pi)^3} \; ,
\ee
in which $\br_{12}\equiv\br_1-\br_2$. This, with the glassy distribution
(83), gives
\be
\label{89}
\rho_G(\br,0) = \int \frac{\ll|\vp_k|^2\gg}{(\om_k+\Dlt)^2} \;
e^{i\bk\cdot\br}\; \frac{d\bk}{(2\pi)^3} \; .
\ee
Taking into account that the main contribution to integral (89) comes
from small $k$, and using the equality
$$
\int_0^\infty \; \frac{x\sin(ax)}{(b^2+x^2)^2} \; dx =
\frac{\pi a}{4b} \; e^{-ab} \; ,
$$
we obtain the glassy density matrix
\be
\label{90}
\rho_G(\br,0) = \rho_G e^{-k_0r} \qquad (k_0 \equiv 2mc) \; .
\ee
This demonstrates that the localized short-range order of the glassy fraction
has the decay length $1/k_0$, which coincides with the healing length.

It is important to stress that the presence of the glassy fraction in the type of systems
under consideration here does not turn the whole system into a Bose glass. This is because
by the commonly accepted classification, the Bose glass phase requires that the
superfluid fraction $n_s$ be zero, which is not the case here.
 Also, the
density of states
$$
\rho(\om) \equiv \frac{4\pi k^2(\om)}{(2\pi)^3}\;
\frac{dk(\om)}{d\om} \; ,
$$
in which $k(\om)$ is defined by the equation $\ep_k=\om$, with $\ep_k$
from Eq. (59) or equivalently Eq. (69), yields
$$
\rho(\om) =
\frac{m^{3/2}\left(\sqrt{\om^2+\Dlt^2}-\Dlt\right )^{1/2}\om}
{\sqrt{2}\; \pi^2\sqrt{\om^2+\Dlt^2}} \; .
$$
This tends to zero at small $\om$ as
$$
\rho(\om) \simeq \frac{\om^2}{2\pi^2 c^3} \qquad
(\om\ra 0) \; .
$$
Thus, the system does not represent a Bose glass, for which $\rho(0)$
must be finite.

To conclude, the action of external random fields on the Bose system induces the
appearance in the latter of the glassy fraction but need not
transform the system as a whole into the Bose glass phase.

\section{Thermodynamic Stability}

It is interesting to study the influence of random potentials on the
thermodynamic stability condition (22). For the Lagrange parameter $\mu_0$
of the condensate fraction, introduced in Eq. (20), we have from the first
of Eqs. (22)
\be
\label{91}
\mu_0 = (\rho +\rho_1 +\sgm_1) \Phi_0 + \mu_G \; ,
\ee
The last term
\be
\label{92}
\mu_G \equiv \frac{1}{2\sqrt{N_0}\; V} \; \sum_{k\neq 0}
<a_k^\dgr \xi_k + \xi_k^* a_k >
\ee
is caused by the direct action of the random potential. From the second
of Eqs. (22), we find
\be
\label{93}
\Phi_0 \;  >  \; \frac{\mu_G}{2\rho_0} \; .
\ee
Thus, the stability condition (93) for the particle interaction strength
$\Phi_0$ of Eq. (2)
depends on the value $\mu_G$.

Equations (72) and (73) show that
$$
<a_k^\dgr \xi_k> \; = \; \ll \al_k^* \xi_k\gg \; = \; -\;
\frac{\ll \vp_k^*\xi_k\gg}{\om_k+ \Dlt} \; .
$$
Thus, the glassy term (92) takes the form
\be
\label{94}
\mu_G = -\; \frac{1}{2\sqrt{N_0}\; V} \; \sum_{k\neq 0} \;
\frac{\ll\vp_k^*\xi_k + \xi_k^*\vp_k\gg}{\om_k+\Dlt} \; .
\ee
One has to exercise considerable caution when analyzing Eq. (94). To stress
this, let us start with the attempt of calculating $\mu_G$ by means of
perturbation theory with respect to weak disorder. Under asymptotically
weak disorder, the limiting approximate solution of Eq. (75) is
$$
\vp_k \simeq \frac{\sqrt{N_0}}{V} \; \xi_k \; .
$$
Substituting this into Eq. (94) yields the perturbative expression
$$
\mu_G' = -\; \frac{1}{V^2}\; \sum_{k\neq 0} \;
\frac{\ll|\xi_k|^2\gg}{\om_k+\Dlt}\; .
$$
With the definition of $R_k$ in (7), we get
$$
\mu_G' = - \int \frac{R_k}{\om_k+\Dlt}\; \frac{d\bk}{(2\pi)^3} \; ,
$$
which is exactly the form obtained in Ref. [24]. Since the correlation
function $R_k$ is assumed to be positive, one has $\mu_G'<0$. Then condition
(93) tells us that the action of the random potential stabilizes the system,
which does not appear plausible, physically, however.

On the other hand, if one interprets the random potential as being caused
by the presence of randomly distributed impurities, which then justifies the
use of analytic regularization procedures for physical integrals, and if one
takes the limit of the uncorrelated spatial white noise potential, defined
in Eq. (9), then one gets from Eq. (94) the different perturbative value
$$
\mu_G'' = \frac{2}{\pi} \; m^2 c R_0 \; .
$$
The latter is positive, contrary to $\mu_G'<0$. In this way, the sign of
the glassy term (94), in a perturbative evaluation, is not independent of
the method of calculation. In other words,
it remains unclear whether the random potential stabilizes or rather
destabilizes the system. This gives a strong hint that the application of perturbation
theory with respect to weak disorder may not be justified for the considered case.
This would show up via inconsistencies, such as divergencies, when going to higher order
in the perturbative calculations we sketched here.

Fortunately, we are able to calculate Eq. (94) without resorting to the
weak-disorder approximation, but by considering instead the whole Eq. (75) exactly.
We immediately obtain then
\be
\label{95}
\mu_G = \frac{\ll\vp_0^*+\vp_0\gg}{2\sqrt{N_0}} \; - \;
\frac{\ll\xi_0^*+\xi_0\gg}{2V} \; = \; 0 \; .
\ee
Thus, we find $\mu_G\equiv 0$ for any type of the random potential and any
strength of disorder. So, the stability condition (93) acquires the simple
form $\Phi_0>0$.

This result teaches us that the action of random potentials on Bose systems
may lead to nonperturbative effects, when calculations for asymptotically
weak disorder can yield incorrect conclusions.

\section{Energy Contribution}

The direct contribution of the random fields to the internal
energy of the system is given by the average of term (65) entering the Hamiltonian
(64), that is, by
\be
\label{96}
E_\xi \; \equiv \; < H_\xi> \; = \; \ll H_\xi \gg \; .
\ee
With relation (73), the latter gives
\be
\label{97}
E_\xi = - \sum_{k\neq 0} \;
\frac{\ll|\vp_k|^2\gg}{\om_k+\Dlt}\; - \;
\frac{1}{V} \sum_{k,p(\neq0)} \;
\frac{\ll\vp_k^* \vp_p\xi_{k-p}\gg}{(\om_k+\Dlt)(\om_p+\Dlt)} \; .
\ee
Exercising now the required caution when dealing with random fields, we
shall not use perturbation theory for weak disorder, but shall instead take
into account the exact Eq. (75) and use the properties $\xi_k^*=\xi_{-k}$,
$\xi_0^*=\xi_0$ in line with Eqs. (6). Employing once Eq. (75), we have
$$
\frac{1}{V} \sum_{p\neq 0} \ll \al_k^* \al_p\xi_{k-p}\gg \; = \;
\frac{\sqrt{N_0}}{V} \; \frac{\ll\vp_k^*\xi_k\gg}{\om_k+\Dlt}\; - \;
\frac{\ll|\vp_k|^2\gg}{\om_k+\Dlt} \; .
$$
This allows us to transform Eq. (97) into
\be
\label{98}
E_\xi = -\; \frac{\sqrt{N_0}}{V} \sum_{k\neq 0} \;
\frac{\ll \vp_k^*\xi_k\gg}{\om_k+\Dlt} \; .
\ee
Invoking once more Eq. (75) in the form
$$
\frac{1}{V} \sum_{p\neq 0} \; \frac{\xi_p^*\vp_p}{\om_k+\Dlt} =
\frac{\sqrt{N_0}}{V}\; \xi_0 \; - \; \vp_0 \; ,
$$
we reduce Eq. (98) to
\be
\label{99}
E_\xi = \sqrt{N_0} \ll \vp_0\gg - \rho_0\ll \xi_0\gg \; ,
\ee
which results in
\be
\label{100}
E_\xi = 0
\ee
for any kind of the random potentials and any strength of disorder.

It is instructive to stress again that the usage of perturbation
theory with respect to weak disorder is not appropriate here. Really,
if we substitute the approximate solution $\vp_k\simeq\sqrt{\rho_0/V}
\xi_k$, corresponding to weak disorder, into Eq. (98), we get the
perturbative energy
$$
E_\xi' = - N \int \frac{n_0R_k}{\om_k+\Dlt}\; \frac{d\bk}{(2\pi)^3}
$$
in the same form as has been obtained by all other authors using
the weak-disorder limit. This result would seem to tell us that the presence of random
potentials diminishes the internal energy.

However, if we interpret the presence of the random potential as the
existence of randomly distributed scatterers, use the analytic regularization
of integrals, and treat the case of white noise, then we find
$$
E_\xi''  = \frac{2}{\pi}\; N m^2 c n_0 R_0 \; .
$$
Hence, the internal energy would now seem to increase with $R_0$. However, both mutually
conflicting perturbative results are at variance with the exact value
(100), which is always zero.

Again, as in the previous Sec. VIII, we come to the conclusion that
perturbation theory with respect to weak disorder can lead to incorrect
results.

\section{Uncondensed Particles}

The properties of uncondensed particles are characterized, first of all,
by their density $\rho_1$ of Eq. (43) and the anomalous average $\sigma_1$
of Eq. (44). Using Eq. (79), the density $\rho_1$ can be presented as the
sum
\be
\label{101}
\rho_1 = \int n_k \; \frac{d\bk}{(2\pi)^3} = \rho_N +
\rho_G
\ee
of the {\it normal density}
\be
\label{102}
\rho_N = \frac{1}{2} \int \left [ \frac{\om_k}{\ep_k} \; {\rm coth}
\left ( \frac{\ep_k}{2T}\right ) - 1 \right ] \; \frac{d\bk}{(2\pi)^3}
\ee
and of the {\it glassy density}
\be
\label{103}
\rho_G \equiv \int n_G(\bk) \; \frac{d\bk}{(2\pi)^3} \; ,
\ee
which can be written as in Eq. (85). The normal density (102) can be
represented as
\be
\label{104}
\rho_N = \frac{(mc)^3}{3\pi^2} \left\{ 1 + \frac{3}{2\sqrt{2}} \;
\int_0^\infty \; \left ( \sqrt{1+x^2}-1\right )^{1/2} \left [
{\rm coth}\left ( \frac{mc^2}{2T}\; x\right ) -1 \right ]\; dx
\right \} \; .
\ee

With the help of  Eq. (80), the anomalous average $\sigma_1$, whose
absolute value is the density of paired particles, can similarly be
written as the sum
\be
\label{105}
\sgm_1 = \int \sgm_k \; \frac{d\bk}{(2\pi)^3} = \sgm_N +
\rho_G
\ee
of two terms. The first term is
\be
\label{106}
\sgm_N = -\; \frac{1}{2} \int \frac{\Dlt}{\ep_k}\;
{\rm coth}\left ( \frac{\ep_k}{2T}\right ) \;
\frac{d\bk}{(2\pi)^3} \; ,
\ee
while the second term is the same glassy density (103). Equation (106)
can be rewritten as
\be
\label{107}
\sgm_N = \sgm_0 \; -\; \frac{1}{2} \int \frac{\Dlt}{\ep_k}
\left [ {\rm coth} \left ( \frac{\ep_k}{2T} \right ) - 1 \right ]
\frac{d\bk}{(2\pi)^3} \; ,
\ee
where
\be
\label{108}
\sgm_0 \; \equiv \; -\; \frac{\Dlt}{2}
\int \frac{1}{\ep_k}\; \frac{d\bk}{(2\pi)^3} \; .
\ee
The integral in Eq. (108) is ultraviolet divergent. This divergence is
well known to be unphysical, since it is caused by the usage of the contact
interaction potential. A general way of treating such integrals is as
follows. First, one restricts to asymptotically weak coupling and applies
the technique of dimensional regularization , which is an accurately
defined mathematical procedure in that limit [5]. Then one analytically
continues the result to finite coupling. The dimensional regularization
gives
$$
\int \frac{1}{\ep_k}\; \frac{d\bk}{(2\pi)^3} = -\;
\frac{2m}{\pi^2}\; \sqrt{m\rho_0\Phi_0} \; .
$$
In this way, we find  for Eq. (108)
\be
\label{109}
\sgm_0 =  \frac{(mc)^2}{\pi^2}\; \sqrt{m\rho_0\Phi_0} \; .
\ee
Changing the variables of integration, Eq. (107) can be represented in
the form
\be
\label{110}
\sgm_N = \sgm_0 \; - \; \frac{(mc)^3}{2\sqrt{2}\;\pi^2} \;
\int_0^\infty \; \frac{(\sqrt{1+x^2}-1)^{1/2}}{\sqrt{1+x^2}} \;
\left [{\rm coth}\left ( \frac{mc^2}{2T}\; x\right ) -1 \right ] \;
dx \; .
\ee

At low temperatures, when $T/mc^2\ll 1$,
Eq. (104) gives
\be
\label{111}
\rho_N \simeq \frac{(mc)^3}{3\pi^2}  + \frac{(mc)^3}{12}
\left ( \frac{T}{mc^2} \right )^2
\ee
and Eq. (110) yields
\be
\label{112}
\sgm_N \simeq \sgm_0 \; - \; \frac{(mc)^3}{12}\left (
\frac{T}{mc^2}\right )^2 \; .
\ee

In the case of weak interactions, such that $mc^2/T_c\ll 1$,
where $T_c$ is the critical temperature
\be
\label{113}
T_c \equiv \frac{2\pi}{m}\left [ \frac{\rho}{\zeta(3/2)}
\right ]^{2/3} \; ,
\ee
Eqs. (104) and (110) lead to
\be
\label{114}
\rho_N \simeq \rho \left ( \frac{T}{T_c}\right )^{3/2} +
\frac{(mc)^3}{3\pi^2}
\ee
and, respectively, to
\be
\label{115}
\sgm_N \simeq \sgm_0 \; - \;
\frac{m^2 cT}{2\pi} \; .
\ee
The analysis of the behavior of $\rho_N$ and $\sgm_N$ shows that
these quantities are characteristic of the Bose system without disorder,
while the explicit influence of the random potential is contained in the
glassy density (103).

\section{Superfluid Fraction}

By a general definition, the superfluid density is the partial
density appearing as a response to a velocity boost,
\be
\label{116}
\rho_s \equiv \frac{1}{3mV}\; \lim_{v\ra 0} \;
\frac{\prt}{\prt{\bf v}}\; \cdot  <\hat\bP_v >_v \; ,
\ee
where the average of the system momentum $\hat\bP_v =\hat\bP+N m\bv$
is calculated with the Hamiltonian
$$
H_v = H + \int \hat\psi^\dgr(\br)\left ( -i\bv \cdot \nabla +
\frac{mv^2}{2} \right ) \hat\psi(\br)\; d\br
$$
of the liquid moving with velocity $\bv$.

The dimensionless superfluid fraction can be represented as
\be
\label{117}
n_s \equiv \frac{\rho_s}{\rho} =  1 \; - \; \frac{2Q}{3T} \; ,
\ee
where $Q$ is the dissipated heat, having for an equilibrium system the
form
\be
\label{118}
Q \equiv \frac{<\hat\bP^2>}{2mN} \; .
\ee
A detailed derivation of Eqs. (117) and (118) can be found, e.g., in
Ref. [4].

Passing to the Fourier transforms, we have
$<\hat\bP^2>=\sum_{k,p}(\bk \cdot \bp)<\hat n_k\hat n_p>$,
where $\hat n_k \equiv a_k^\dgr a_k$. In the HFB approximation,
$$
<\hat n_k\hat n_p>\; = \; n_k n_p + \dlt_{kp} n_k (1+ n_k) +
\dlt_{-kp} \sgm_k^2 \; .
$$
Then Eq. (118) assumes the form
\be
\label{119}
Q = \frac{1}{\rho} \int \frac{k^2}{2m} \left ( n_k + n_k^2 -
\sgm_k^2\right ) \frac{dk}{(2\pi)^3} \; .
\ee
Taking into account Eqs. (79) and (80), we may represent the dissipated
heat (119) as the sum
\be
\label{120}
Q = Q_N + Q_G
\ee
of two terms. Here the first term
\be
\label{121}
Q_N = \frac{1}{8m\rho} \int \frac{k^2}{{\rm sinh}^2(\ep_k/2T)} \;
\frac{d\bk}{(2\pi)^3}
\ee
is the heat dissipated by normal uncondensed particles. And the second
term
\be
\label{122}
Q_G = \frac{1}{2m\rho} \int \frac{k^2\ll|\vp_k|^2\gg}{\ep_k(\om_k+\Dlt)} \;
{\rm coth}\left ( \frac{\ep_k}{2T}\right ) \; \frac{d\bk}{(2\pi)^3}
\ee
is the heat dissipated by the glassy fraction.

Equation (121) can be rewritten as
\be
\label{123}
Q_N = \frac{(mc)^5}{4\sqrt{2}\;\pi^2 m\rho}\; \int_0^\infty \;
\frac{(\sqrt{1+x^2}-1)^{3/2} x\; dx}
{\sqrt{1+x^2}\;{\rm sinh}^2(mc^2x/2T)} \; .
\ee
At low temperatures, such that $T/mc^2\ll 1$, we get
\be
\label{124}
Q_N \simeq \frac{\pi^2(mc)^5}{15m\rho} \left (
\frac{T}{mc^2}\right )^5 \; .
\ee
And in the limit of weak interactions, when $mc^2/T_c\ll 1$, we find
\be
\label{125}
Q_N \simeq \frac{3}{2}\; T \left [ \left ( \frac{T}{T_c}\right )^{3/2}
- \; \frac{\zeta(1/2)}{\zeta(3/2)} \left ( \frac{T}{T_c}\right )^{1/2}
\frac{mc^2}{T_c} \right ] \; ,
\ee
where $\zeta(\cdot)$ is a Riemann zeta function.

\section{Sound Velocity}

The sound velocity $c$ enters in the majority of the above expressions.
The velocity itself is defined through Eq. (71), which can be written as
\be
\label{126}
mc^2 =  (\rho - \rho_1 +\sgm_1) \Phi_0 \; ,
\ee
taking into account that $\rho_0=\rho-\rho_1$. According to Eqs. (101)
and (105), we have
\be
\label{127}
\rho_1 = \rho_N + \rho_G \; , \qquad \sgm_1 =\sgm_N +\rho_G \; .
\ee
Therefore Eq. (126) becomes
\be
\label{128}
mc^2 = (\rho -\rho_N +\sgm_N) \Phi_0 \; .
\ee

It is convenient to work with the dimensionless fractions
\be
\label{129}
n_N \equiv \frac{\rho_N}{\rho} \; , \qquad
\sgm \equiv \frac{\sgm_N}{\rho} \; .
\ee
Since $n_0+n_1=1$ and $n_1=n_N+n_G$, the normalization
\be
\label{130}
n_0 + n_N + n_G  = 1
\ee
holds true.

Let us define the gas parameter
\be
\label{131}
\gm \equiv \rho^{1/3} a_s \; .
\ee
and the dimensionless sound velocity
\be
\label{132}
s \equiv \frac{mc}{\rho^{1/3}} \; .
\ee
Then, taking into consideration the interaction strength (2), equation
(128) for the sound velocity can be reduced to the dimensionless form
\be
\label{133}
s^2 =  4\pi\gm ( 1 - n_N + \sgm) \; .
\ee

At first glance it might seem that the sound velocity, being the solution
of Eq. (133), does not depend on the glassy fraction $n_G$ induced by the
random fields. That fraction is defined by Eqs. (85) and (87) which give
combined
\be
\label{134}
n_G = \frac{1}{\rho} \int \frac{\ll|\vp_k|^2\gg}{(\om_k+\Dlt)^2} \;
\frac{d\bk}{(2\pi)^3} \; .
\ee
However, through normalization (130), $n_G$ influences the condensate fraction
$n_0$, and the latter enters the anomalous fraction $\sgm$, thus, influencing
the sound velocity through Eq. (133). For example, at zero temperature,
according to Eqs. (104) to (112), we have
\be
\label{135}
n_N = \frac{s^3}{3\pi^2} \; , \qquad \sgm = \frac{\sgm_0}{\rho} =
\frac{2s^2}{\pi^{3/2}}\; \sqrt{\gm n_0} \; , \qquad
n_0 = 1 \; - \; \frac{s^3}{3\pi^2}\; - \; n_G \qquad (T=0) \; .
\ee
Increasing disorder increases the glassy fraction $n_G$, so, decreases
the condensate fraction $n_0$, which decreases $\sgm$. At the same time,
the normal fraction $n_N$ also decreases. Since $n_N$ and $\sgm$ enter
Eq. (133) with opposite signs, their changes almost compensate each other.
Numerical calculations show that the sound velocity $s$ as a function of
the disorder strength slightly decreases with the latter.

\section{Structure Factor}

The structure factor of a random system is defined as the stochastic average
\be
\label{136}
S(\bk) \equiv \; \ll S_H(\bk) \gg
\ee
of the frozen factor
\be
\label{137}
S_H(\bk) = \frac{1}{N} \; \int \left [ <\hat n(\br)\hat n(\br') >_H \; -
\; <\hat n(\br)>_H <\hat n(\br')>_H \right ] \;
e^{-i\bk \cdot(\br-\br')} d\br d\br' \; ,
\ee
expressed through the quantum averages, in which
$\hat n(\br)\equiv\hat\psi^\dgr(\br)\hat\psi(\br)$. Note that Eq. (136),
in the theory of random systems, is called the connected structure factor.
With the Fourier transform
\be
\label{138}
\hat\rho_k = \int \hat n(\br) \; e^{-i\bk\cdot\br} \; d\br \; ,
\ee
Eq. (137) becomes
\be
\label{139}
S_H(\bk) = \frac{1}{N} \left ( <\hat\rho_k^+ \hat\rho_k>_H \; - \;
<\hat\rho^+_k>_H <\hat\rho_k>_H\right ) \; .
\ee
Invoking the Bogolubov shift (13), for Eq. (138), we have
\be
\label{140}
\hat\rho_k = \dlt_{k0} N_0 + \sum_{p\neq 0} a_{k+p}^\dgr a_p \; +
\; \sqrt{N_0}\left ( a_{-k}^\dgr + a_k\right ) \; .
\ee
The quantum averaging of Eq. (140) gives
\be
\label{141}
<\hat\rho_k>_H \; = \dlt_{k_0} N_0 +
\sum_{p\neq 0} <a_{k+p}^\dgr a_p>_H \; + \; \sqrt{N_0}\; \left (
\al_{-k}^* + \al_k \right ) \; ,
\ee
where $\al_k$ is defined in Eq. (47). Calculating the first term in Eq.
(139), we arrange the operator product in the normal form and use the
second-order procedure, following the standard calculations, the same
as for Bose systems without disorder [7,35]. Then for the structure
factor (136), we find
\be
\label{142}
S(\bk) = 1 + 2 (n_k +\sgm_k) - 4\ll |\al_k|^2 \gg \; .
\ee
Substituting here Eqs. (79) and (80), we obtain
\be
\label{143}
S(\bk) = \frac{k^2}{2m\ep_k} \; {\rm coth}\left (
\frac{\ep_k}{2T}\right ) \; .
\ee
The central value of the structure factor is known to be related to
the isothermal compressibility
\be
\label{144}
\kappa_T \equiv -\; \frac{1}{V} \left ( \frac{\prt V}{\prt P}
\right )_T = \frac{1}{\rho} \left ( \frac{\prt\rho}{\prt P}
\right )_T = \frac{S(0)}{\rho T} \; ,
\ee
where $P$ is pressure. To emphasize the role of the glassy fraction,
the central structural factor can be written as
\be
\label{145}
S(0) = \frac{T}{mc_0^2} \; + \; An_G \; ,
\ee
where $A\equiv-2Tc_0'/mc_0^3$; $c_0$ is the sound velocity in a system
without disorder, and $c_0'\equiv\prt c/\prt n_G$ at the value $n_G=0$.
From numerical calculations it follows that the coefficient of $A$ is
positive. Thus, the above expressions show that the random field, via
inducing the glassy fraction $n_G$, leads to an increase of the density
fluctuations, the isothermal compressibility, and the structure factor.
The physics of these results seems to be clear. An additional external
random potential should lead to the increased scattering of either
light or neutrons, which is characterized by an increase of the structure
factor.

\section{White Noise}

The influence of the random potential on physical characteristics comes
through the correlator $\ll|\vp_k|^2\gg$. To calculate the latter explicitly,
we need, first, to solve the random-field equation (75) and, second, to
specify the type of the random potential, which till now has been arbitrary.

Let us consider Eq. (75) assuming that in the sum of its second term the
main contribution comes from the region of small momenta (see Appendix A),
so that this equation can be represented as
\be
\label{146}
\vp_k = \frac{\sqrt{N_0}}{V}\; \xi_k \; - \;
\frac{1}{V} \sum_{p\neq 0} \; \frac{\xi_k\vp_p}{\om_p+\Dlt} \; .
\ee
This is the Fredholm equation of the second kind with a separable kernel.
Such an equation can be solved exactly. The corresponding exact solution
is
\be
\label{147}
\vp_k = \frac{\sqrt{N_0}}{V}\;
\frac{\xi_k}{1+\frac{1}{V}\sum_{p\neq 0}
\frac{\xi_p}{\om_p+\Dlt}} \; .
\ee
Calculating $\ll|\vp_k|^2\gg$ with Eq. (147), we can use the expansion
$$
\frac{1}{(1-x)^2} = \sum_{n=0}^\infty (n+1) x^n \; ,
$$
which requires the knowledge of the stochastic correlators such as
$\ll\xi_{k_1}\xi_{k_2}\ldots\xi_{k_n}\gg$.

To define these correlators explicitely, we consider the case of the
Gaussian white noise [49]. Then we obtain
\be
\label{148}
\ll |\vp_k|^2\gg\; = \; \rho_0 R_0 \sum_{n=0}^\infty (2n+1)!!
\left ( \frac{mR_0}{4\pi c}\right )^n \; ,
\ee
where the integral
$$
\int \frac{1}{(\om_k + \Dlt)^2} \; \frac{d\bk}{(2\pi)^3} =
\frac{m}{4\pi c}
$$
has been used. Note that the right-hand side of Eq. (148) does not
depend on $k$. This allows us to find the explicit expression for
the glassy density (85), which becomes
\be
\label{149}
\rho_G = \frac{m}{4\pi c} \ll |\vp_k|^2\gg \; ,
\ee
where the right-hand side is given by the series (148).

As is clear from its form, series (148) is asymptotic with respect to the
parameter $mR_0/4\pi c$. In order to define the quantity $\ll|\vp_k|^2\gg$
for finite values of the latter parameter, it is necessary to employ a
resummation procedure for series (148). For example, we could resort to the
Pad\'e summation [50]. Here we shall use another, more general and accurate
method, based on the self-similar approximation theory [51--53]. We shall
make use of the method of self-similar factor approximants [54--56]. This
method was shown to be more general than that of Pad\'e approximants and,
contrary to the latter, being uniquely defined. The method we use is 
sketched in the Appendix B.

For convenience, we introduce the dimensionless {\it noise parameter}
\be
\label{150}
\nu \equiv \frac{7m^2R_0}{4\pi\rho^{1/3}} \; .
\ee
Then, representing the sum of series (148) by the self-similar factor
approximant of second order, we obtain for the glassy density (149) the
expression
\be
\label{151}
\rho_G = \frac{\rho_0\nu}{7s^{4/7}(s-\nu)^{3/7}} \; ,
\ee
in which $s$ is the dimensionless sound velocity (132).

Taking into account normalization (130), we find the condensate fraction
\be
\label{152}
n_0 = \frac{7s^{4/7}(s-\nu)^{3/7}}{\nu+7s^{4/7}(s-\nu)^{3/7}}\;
(1-n_N)
\ee
and the glassy fraction
\be
\label{153}
n_G = \frac{\nu(1-n_N)}{\nu+7s^{4/7}(s-\nu)^{3/7}} \; ,
\ee
which are expressed through the normal fraction $n_N\equiv\rho_N/\rho$,
with $\rho_N$ given by Eq. (104).

The case of weak disorder corresponds to a small noise parameter (150).
Then the condensate fraction (152) is
\be
\label{154}
n_0 \simeq \left ( 1\; - \; \frac{\nu}{7s} \; - \;
\frac{2\nu^2}{49s^2} \right ) (1 -n_N)
\ee
and the glassy fraction (153) becomes
\be
\label{155}
n_G \simeq \frac{\nu}{7s} \left ( 1 + \frac{2\nu}{7s}
\right ) (1-n_N) \; ,
\ee
when $\nu\ll 1$.

If, in addition, atomic interactions are asymptotically weak, such that
$a_s\ra 0$, then the glassy fraction (155) tends to
\be
\label{156}
n_G \simeq \frac{\nu}{7s}\; n_0 = \frac{mn_0R_0}{4\pi c} \; .
\ee
In this limit, the sound velocity acquires the Bogolubov form
\be
\label{157}
c \simeq \sqrt{\frac{\rho_0\Phi_0}{m}} =
\frac{2}{m}\; \sqrt{\pi\rho a_s} \; .
\ee
As a result, the glassy fraction (156) transforms to
\be
\label{158}
n_G \simeq \frac{m^2R_0}{8\pi^{3/2}}\;
\sqrt{\frac{n_0}{\rho a_s}} \; ,
\ee
which exactly coincides with the expression found by Huang and Meng
[22] in the limit of asymptotically weak interactions and weak disorder.

We may notice that the noise parameter $\nu$ enters Eqs. (151) to (155)
in the combination
$$
\frac{\nu}{s} = \frac{7mR_0}{4\pi c} \; .
$$
It would, therefore, be tempting to consider the ratio $\nu/s$ as a
new parameter. However, this ratio becomes really a parameter solely for
asymptotically weak interactions, when
$$
\frac{\nu}{s} \simeq \frac{7m^2R_0}{8\pi^{3/2}\sqrt{\rho a_s}} \qquad
(a_s \ra 0) \; .
$$
But at finite interactions, the sound velocity $c=c(T,\rho,a_s)$ is
a complicated function of temperature, density, and scattering length.
Respectively, the dimensionless sound velocity $s=s(T,\rho,\gm)$ is a
function of temperature, density, and the gas parameter, defined by Eq.
(133). Hence, at finite interactions, temperatures, and disorder strength,
the situation is more involved and one cannot reduce the consideration to
dealing with the ratio $\nu/s$, which is not anymore a parameter.

Equations (152) and (153) show that when atomic interactions are
switched off, so that $s\ra 0$, then there are no positive solutions for
the fractions $n_0$ and $n_G$ for any finite noise parameter $\nu$. This
means that the ideal Bose-condensed gas is {\it stochastically unstable},
in the sense that any infinitesimally weak disorder completely destroys
the Bose-Einstein condensate, rendering the system to the normal state.

In the case of an interacting Bose-condensed system with a finite gas
parameter $\gm$, the system is stable below a critical noise parameter
$\nu_c=\nu_c(T,\rho,\gm)$. Increasing $\nu$ diminishes the condensate
fraction but increases the glassy fraction. Reaching the critical value
$\nu_c$, the system undergoes a {\it first-order phase transition}, when
$n_0$ and $n_G$ jump to zero, after which the normal phase prevails with
$n_N=1$. This is in agreement with a first-order phase transition found
in the particular case of zero temperature and asymptotically weak
interactions [28]. According to our numerical estimates, the jumps of
$n_0$ and $n_s$ are close to those found in Ref. [28] at the transition
point $\nu_c$.

When disorder is absent, the system displays the second-order phase
transition at the critical temperature (113) coinciding with that of
the ideal Bose gas, which follows from expansions (114) and (125). As
soon as there appears disorder, with any finite noise parameter $\nu$,
the phase transition becomes of first order. At asymptotically small
$\nu\ra 0$, numerical estimates give the shift of the critical
temperature $\dlt T_c\sim-2\nu/9\pi$, which is close to the shift
found in Refs. [24,25].

To analyze the behavior of the superfluid fraction (117), we need to
know the dissipated heat (120). The part of this quantity, due to normal
particles, is given by Eq. (123). Another part, caused by the heat
dispersed by the glassy fraction, is defined by Eq. (122). With the
white-noise relation (149), expression (122) can be represented as
\be
\label{159}
Q_G = \frac{8}{\pi}\; m c^2 n_G I \left (
\frac{mc^2}{T} \right ) \; ,
\ee
where the notation
\be
\label{160}
I(\ep) \equiv \int_0^\infty {\rm coth}\left ( \ep x\; \sqrt{1+x^2}
\right ) \frac{x^3\; dx}{(1+x^2)^{3/2}}
\ee
is introduced.

Integral (160) diverges for any finite $\ep$, so that one has to invoke
some regularization of this integral. There are several ways to regularize
the integral, all of which yield the same result.

First of all, we understand that the divergence of the above integral
is caused by the white noise. For a colored noise, we should go back
to Eq. (122), in which $\ll|\vp_k|^2\gg$ would be a diminishing function
of $k$, but not a constant in $k$, as for the white noise in Eq. (148).
Then integral (122) would be convergent. For such a colored noise, we
could approximate Eq. (122) as
$$
\frac{1}{2m\rho}\; \int \;
\frac{k^2\ll|\vp_k|^2\gg}{\ep_k(\om_k+\Dlt)}\;
{\rm coth}\left ( \frac{\ep_k}{2T}\right ) \frac{d\bk}{(2\pi)^3}\;
\cong \;  \frac{T}{m\rho} \; \int \;
\frac{k^2\ll|\vp_k|^2\gg}{\ep_k^2(\om_k+\Dlt)}\;
\frac{d\bk}{(2\pi)^3} \; .
$$
Passing after this to the white-noise relation (149), we obtain
\be
\label{161}
Q_G =  2 n_G T \; .
\ee

Another way of regularizing Eq. (160) is a kind of analytic
regularization, where one, first, takes the integral for that region of
the parameter $\ep$, where the integral converges, and then analytically
continues the result to arbitrary values of this parameter. The sole
domain of $\ep$, where integral (160) converges, is the region of
asymptotically small $\ep\ra 0$. Then
\be
\label{162}
I(\ep) \simeq \frac{1}{\ep} \; \int_0^\infty \;
\frac{x^2\; dx}{(1+x^2)^2} = \frac{\pi}{4\ep} \; .
\ee
Substituting this into Eq. (159) gives again Eq. (161).

We can also apply for integral (160) a resummation regularization, when
the integral, first, is represented as a series, after which the series is
reorganized with the help of a resummation procedure. Two variants of the
self-similar regularization are described in the Appendix C, both of which
lead to the same answer (161) as two previous regularizations considered
above. It is important to stress that we have accomplished several ways
of regularizing integral (160) in order to prove that the result does
not depend on the regularization procedure involved.

Thus, combining Eqs. (117), (120), and (161), we find the expression for
the superfluid fraction
\be
\label{163}
n_s  = 1\; - \; \frac{4}{3}\; n_G \; - \; \frac{2Q_N}{3T} \; ,
\ee
in which $n_G$ is defined in Eq. (153) and $Q_N$, in Eq. (123). It is
worth stressing that, though the superfluid fraction (163) is linear with
respect to the glassy fraction $n_G$, it is far from being linear with
respect to the strength of disorder $\nu$, as follows from expression (153)
for the glassy fraction.

At low temperatures, when $T\ll mc^2$, the superfluid fraction (163) is
\be
\label{164}
n_s \simeq 1 \; - \;
\frac{4\nu/3}{\nu+7s^{4/7}(s-\nu)^{3/7}} \left [ 1 \; - \;
\frac{s^3}{3\pi^2} \; - \; \frac{s^3}{12} \left (
\frac{T}{mc^2}\right )^2 \right ] \; - \;
\frac{2\pi^2 s^3}{45}\left (\frac{T}{mc^2}\right )^4 \; .
\ee
And in the case of weak interactions, when $mc^2\ll T_c$, the fraction (163)
has the form
$$
n_s \simeq 1 - \left ( \frac{T}{T_c}\right )^{3/2} - \;
\frac{\zeta(1/2)}{\zeta(3/2)} \left ( \frac{T}{T_c}\right )^{1/2}
\frac{mc^2}{T_c} \; -
$$
\be
\label{165}
- \; \frac{4\nu/3}{\nu+7s^{4/7}(s-\nu)^{3/7}} \left [ 1 \; - \;
\frac{s^3}{3\pi^2} \; - \left ( \frac{T}{T_c}\right )^{3/2}
\right ] \; .
\ee
The superfluid fraction $n_s$ can be either larger or smaller than the condensate fraction
$n_0$, depending on temperature, the strength of interactions, that is, on the
gas parameter $\gm$, and on the strength of disorder $\nu$. Increasing $\nu$ leads to
the simultaneous disappearance of the superfluid and condensate fractions at the same critical
$\nu_c$ through a first-order phase transition.  This transition takes place between the
superfluid phase, with $n_0\neq 0$, $n_s\neq 0$, $n_G\neq 0$, and $n_N<1$,
and the normal phase with $n_0=0$, $n_s=0$, $n_G=0$, and $n_N=1$.

\section{Conclusion}

A self-consistent mean-field theory has been developed for Bose systems
in random external potentials. The suggested approach makes it possible
to consider arbitrarily strong interactions and an arbitrary strength of
disorder. In general, the Bose system consists of the following components:
the condensate fraction $n_0$, the normal fraction $n_N$, the glassy fraction
$n_G$, and the superfluid fraction $n_s$. In the limit of asymptotically
weak interactions and disorder, the known results are reproduced. When
increasing the strength of disorder, a first-order phase transition occurs
from the superfluid phase to the normal phase. For the class of models
we considered we have found no pure Bose glass phase. The temperature
for the occurrence of the first-order phase transition turns out to be
lower than the critical temperature $T_c$ of the second-order phase
transition for a Bose system without disorder. The presence of disorder
slightly lowers the sound velocity, but increases the density fluctuations,
the isothermal compressibility, and the structure factor.

It is interesting that switching on disorder may lead to nonperturbative
effects. For instance, the uniform ideal Bose gas is stochastically unstable with
respect to infinitesimally small noise. Perturbation theory cannot be used
to calculate the internal energy contributed by random fields. Nor is
perturbation theory sufficient when analyzing the stability condition
related to the minimization of the thermodynamic potential.

The aim of present paper has been to develop an approach for considering
Bose systems with any interaction strength and arbitrary strength of
disorder and to describe the general properties of such systems. We have
restricted ourselves to investigating those results that could be derived
by analytic means. The overall quantitative study of the system properties
requires to solve the intricate system of equations for functions of
temperature $T$, density $\rho$, gas parameter $\gm$, and noise parameter
$\nu$. Such an investigation can be accomplished only numerically. In view
of the length of the present paper, we prefer not to overload it further
by these numerical calculations. They will be presented in separate
publications.

\vskip 5mm

{\bf Acknowledgements}

\vskip 3mm

The authors acknowledge financial support from the German Research
Foundation. One of us (V.I.Y.) is grateful to E.P. Yukalova for
discussions and to V. Krakoviack for helpful comments, the other (R.G.) 
wants to thank A. Pelster and P. Navez
for some useful remarks.

\newpage

{\Large{\bf Appendix A}}

\vskip 5mm

Throughout the paper, we have assumed that the external potential $\xi(\br)$
possesses the Fourier transform $\xi_k$ for all $\bk$. The zero-momentum
transform
$$
\xi_0 = \int \xi(\br)\; d\br
$$
exists in the sense that it is finite. This implies that the function
$\xi(\br)$ is integrable, such that
$$
\left | \int \xi(\br)\; d\br \right | < \infty \; .
$$
For a function $\xi(\br)$ integrable in that sense, the Riemann-Lebesgue
theorem (for details see e.g. Refs. [57,58]) states that the related
Fourier transforms $\xi_k$ is finite for all $\bk$ and tends to zero as
$k\equiv|\bk|$ tends to infinity, $|\xi_k|<\infty$,
$$
\xi_k \ra 0 \qquad (k\ra \infty) \; .
$$
How $\xi_k$ tends to zero depends on further properties of
$\xi(\br)$. For instance, if we can use that the function $\xi(\br)$ is
finite at $\br=0$, that is,
$$
|\xi(0)| < \infty \; ,
$$
then we have
$$
\left | \frac{1}{V} \sum_k \xi_k \right | < \infty \; .
$$
Replacing summation by integration in the standard manner, we get
$$
\left | \int \xi_k\; d\bk \right | < \infty \; .
$$
It follows from here that, when $k\ra\infty$, then $\xi_k$ tends to zero at
least as
$$
|\xi_k| \; \propto \; \frac{1}{k^\al} \qquad (\al > 3) \; .
$$
in three or more dimensions. The field $\vp_k$ is related to $\xi_k$ by Eq. (75). Assuming that $\vp_k$
has the same asymptotic behavior as $\xi_k$, so that $\vp_k$ tends to zero
at large $k\ra\infty$ not slower than
$$
|\vp_k| \; \propto \; \frac{1}{k^\al} \qquad (\al> 3) \; ,
$$
we find solution (147), which confirms this assumption. With the field
$\vp_k$ possessing this asymptotic behavior and the form of $\om_k$ defined
in Eq. (68), the ratio $\vp_k/(\om_k+\Dlt)$ tends to zero at large $k$ not
slower than
$$
\left | \frac{\vp_k}{\om_k+\Dlt}\right | \; \propto \;
\frac{1}{k^n} \qquad (n> 5) \; .
$$
This justifies the transformation of Eq. (75) into Eq. (146).

\newpage

{\Large{\bf Appendix B}}

\vskip 5mm

For the summation of series (148), we used the method of self-similar factor 
approximants [54--56], which was shown to be more general and more accurate 
than the method of Pad\'e approximants. In addition, the latter method is 
known to be not uniquely defined, in the sense that for each finite series 
of order $k$, there exists a table of Pad\'e approximants $P_{[M/N]}$, with 
$M+N=k$. Contrary to this, for each finite series of order $k$, there is 
just one factor approximant. This is why we prefer here to use the method
of self-similar factor approximants. The construction of the factor 
approximants is done in the following way.

Assume that for a function $f(x)$, there is an asymptotic expansion at 
$x\ra 0$, so that $f(x)$ can be represented by
$$
f_k(x) = f_0(x) \; \sum_{n=0}^k a_n x^n \; .
$$
Since $a_0$ can always be included in $f_0(x)$, we may set, without the 
loss of generality, that $a_0=1$. The self-similar factor approximant, 
extrapolating the finite series $f_k(x)$ to the region of arbitrary $x$, is 
defined as
$$
f_k^*(x) = f_0(x) \prod_{i=1}^{N_k} ( 1 + A_i x)^{n_i} \; ,
$$
where
\begin{eqnarray}
\nonumber
N_k = \left \{ \begin{array}{ll}
k/2 \; , & k=2,4,\ldots \\
(k+1)/2 \; , & k=3,5,\ldots \end{array} .
\right.
\end{eqnarray}
The coefficients $A_i$ and exponents $n_i$ are given by the equations
$$
\sum_{i=1}^{N_k} n_i A_i^n = B_n \qquad (n=1,2,\ldots, k) \; ,
$$
in which
$$
B_n \equiv \frac{(-1)^{n-1}}{(n-1)!}\; \lim_{x\ra 0} \; 
\frac{d^n}{dx^n}\; \ln \; \frac{f_k(x)}{f_0(x)}
$$
and $A_1$ is set to one for odd $k$. As is evident, the parameters $A_i$, 
$n_i$, and $B_n$, being calculated for each given $k$-th order approximation, 
depend on the approximation order $k$, so that $A_i=A_{ik}$, $n_i=n_{ik}$, and 
$B_n=B_{nk}$. However, for the simplicity of notations, this order dependence 
is not shown explicitly. A detailed description of the method of self-similar 
factor approximants is given in Refs. [54--56].

\newpage

{\Large{\bf Appendix C}}

\vskip 5mm

Integral (160) can be regularized by means of the resummation regularization
based on the self-similar approximation theory [51--53]. The procedure is as
follows. One, first, introduces a cutoff $L$ making the integral always
converging,
$$
I_L(\ep) \equiv \int_0^L {\rm coth}\left (\ep x\; \sqrt{1+x^2}\right )
\frac{x^3\; dx}{(1+x^2)^{3/2}} \; .
$$
Removing this cutoff would return us back to the integral
$$
I(\ep) = \lim_{L\ra\infty} I_L(\ep) \; .
$$
Instead, we may represent the integral $I_L(\ep)$ as a series with the help
of the expansion
$$
{\rm coth} z = \sum_{n=0}^\infty \frac{2^{2n} B_{2n}}{(2n)!}\;
z^{2n-1} \; ,
$$
yielding the series
$$
I_L(\ep) = \frac{1}{\ep} \sum_{n=0}^\infty a_{2n} \ep^{2n} \; ,
$$
with the coefficients
$$
a_{2n} \equiv \frac{2^{2n}B_{2n}}{(2n)!} \int_0^L \;
\frac{x^{2+2n}dx}{(1+x^2)^{2-n}} \; ,
$$
where $B_n$ are Bernoulli numbers. When $L\ra\infty$, then
$$
\lim_{L\ra\infty} a_0 = \frac{\pi}{4}
$$
and other coefficients behave as
$$
a_{2n} \simeq \frac{2^{2n}B_{2n}}{(2n)!(4n-1)}\; L^{4n-1} \; ,
$$
with $n=1,2,\ldots$.

The series for $I_L(\ep)$ can be represented with the help of the
self-similar factor approximants [54--56] as
$$
I_L^*(\ep) = \frac{a_0}{\ep} \prod_{n=1}^\infty \left (
1 + A_n \ep^2\right )^{\al_n} \; ,
$$
where the parameters $A_n$ and powers $\al_n$ are uniquely defined from
the re-expansion procedure, when $I_L^*(\ep)$ is expanded in powers of
$\ep^2$ and compared with the initial series for $I_L(\ep)$. Then all $A_n$
and $\al_n$ are uniquely expressed through the coefficients $a_{2n}$. For
example, in lower orders, we have
$$
A_1 = \frac{b_2^2-2b_4}{b_2} \; , \qquad
\al_1 =\frac{b_2^2}{b_2^2-2b_4} \; ,
$$
where
$$
b_n \equiv \frac{a_n}{a_0} \simeq \frac{2^{2n+n}B_{2n}}{(2n)!(4n-1)\pi}\;
L^{4n-1} \; .
$$
Then
$$
b_2 \simeq \frac{8}{9\pi}\; L^3 \; , \qquad b_4 \simeq -\;
\frac{4}{315\pi}\; L^7 \; ,
$$
etc. Because of this,
$$
A_1 \simeq \frac{1}{35}\; L^4 \; , \qquad \al_1 \simeq \frac{280}{9\pi}\;
\frac{1}{L} \; ,
$$
and so on.

Removing the cutoff in the self-similar approximant $I_L^*(\ep)$ by setting
$L\ra\infty$, we use the limit
$$
\lim_{L\ra\infty}\left ( 1 + L^m\right )^{1/L^n} = 1 \; ,
$$
valid for any $m>0$ and $n>0$. As a result, we have
$$
\lim_{L\ra\infty} I_L^*(\ep) = \frac{\pi}{4\ep} \; .
$$
Substituting this into Eq. (159), we come to the same form of the dispersed
heat (161) as obtained earlier.

Another variant of the resummation regularization would be by summing the
series for $I_L(\ep)$ in the form of the self-similar exponential
approximants [59,60]. This procedure gives
$$
I_L^*(\ep) = \frac{a_0}{\ep} \; \exp\left ( b_2 \ep^2 \;
\exp\left ( b_4 \ep^2 \cdots \right ) \right ) \; ,
$$
where again $b_n\equiv a_n/a_0$. When setting $L\ra\infty$, we use the fact
that the Bernoulli numbers are alternating in sign, so that $b_2$, $b_6$,
$b_{10}$, and so on tend, polynomially in $L$, to plus infinity, while
$b_4$, $b_8$, $b_{12}$, and like that tend polynomially to minus infinity.
Then we obtain
$$
I_\infty^*(\ep) =\frac{a_0}{\ep} = \frac{\pi}{4\ep} \; ,
$$
which again leads to the same form (161). In this way, all considered
variants of regularizing integral (160) give us the same expression (161),
which confirms its general validity.


\newpage

\begin{thebibliography}{99}
\bibitem{1}
G.K.S. Wong, P.A. Crowell, H.A. Cho, and J.D. Reppy, Phys. Rev. Lett. {\bf
65}, 2410 (1990).

\bibitem{2}
J.D. Reppy, J. Low Temp. Phys. {\bf 87}, 205 (1992).

\bibitem{3}
C.J. Pethick and H. Smith, {\it Bose-Einstein Condensation in Dilute Gases}
(Cambridge University Press, Cambridge, 2002).

L. Pitaevskii and S. Stringari, {\it Bose-Einstein Condensation}
(Clarendon, Oxford, 2003).

\bibitem{4}
P.W. Courteille, V.S. Bagnato, and V.I. Yukalov, Laser Phys. {\bf 11},
659 (2001).

\bibitem{5}
J.O. Andersen, Rev. Mod. Phys. {\bf 76}, 599 (2004).

\bibitem{6}
K. Bongs and K. Sengstock, Rep. Prog. Phys. {\bf 67}, 907 (2004).

\bibitem{7}
V.I. Yukalov, Laser Phys. Lett. {\bf 1}, 435 (2004).

\bibitem{8}
D. Clement {\it et al}., Phys. Rev. Lett. {\bf 95}, 170409 (2005).

\bibitem{9}
C. Fort {\it et al}., Phys. Rev. Lett. {\bf 95}, 170410 (2005).

\bibitem{10}
H. Gimperlein, S. Wessel, J. Schmiedmayer, and L. Santos, Phys. Rev. Lett.
{\bf 95}, 170401 (2005).

\bibitem{11}
L. Fallani, J.E. Lye, V. Guarrera, C. Fort, and M. Inguscio, e-print
cond-mat/0603655 (2006).

\bibitem{12}
J.A. Hettz, L. Fleishman, and P.W. Anderson, Phys. Rev. Lett. {\bf 43}, 942
(1979).

\bibitem{13}
M.P.A. Fisher, P.B. Weichman, G. Grinstein, and D.S. Fisher, Phys. Rev. B
{\bf 40}, 546 (1989).

\bibitem{14}
R. Pugatch, N. Bar-Gill, N. Katz, E. Rowen, and N. Davidson, e-print
cond-mat/0603571 (2006).

\bibitem{15}
R.T. Scalettar, G.G. Batrouni, and G.T. Zimanyi, Phys. Rev. Lett. {\bf 66},
3144 (1991).

\bibitem{16}
K.G. Singh and D.S. Rokhsar, Phys. Rev. B {\bf 46}, 3002 (1992).

\bibitem{17}
K.G. Singh and D.S. Rokhsar, Phys. Rev. B {\bf 49}, 9013 (1994).

\bibitem{18}
A. De Martino, M. Thorwart, R. Egger, and R. Graham, Phys. Rev. Lett. {\bf
94}, 060402 (2005).

\bibitem{19}
P. Hitchcock and E.S. S\o rensen, e-print cond-mat/0601180 (2006).

\bibitem{20}
P. Hitchcock and E.S. S\o rensen, e-print cond-mat/0601182 (2006).

\bibitem{21}
K.V. Krutitsky, A. Pelster, and R. Graham, New J. Phys. {\bf 8}, 187 (2006).

\bibitem{22}
H. Huang and H.F. Meng, Phys. Rev. Lett. {\bf 69}, 644 (1992).

\bibitem{23}
S. Giorgini, L. Pitaevskii, and S. Stringari, Phys. Rev. B {\bf 49}, 12938
(1994).

\bibitem{24}
A.V. Lopatin and V.M. Vinokur, Phys. Rev. Lett. {\bf 88}, 235503 (2002).

\bibitem{25}
O. Zobay, Phys. Rev. A {\bf 73}, 023616 (2006).

\bibitem{26}
R. Graham and A. Pelster, e-print cond-mat/0508306 (2005).

\bibitem{27}
G.E. Astrakharchik, J. Boronat, J. Casulleras, and S. Giorgini, Phys. Rev. A
{\bf 66}, 023603 (2002).

\bibitem{28}
P. Navez, A. Pelster, and R. Graham, Appl. Phys. B, in press (2006).

\bibitem{29}
F.A. Berezin, {\it Method of Second Quantization} (Academic, New York 1966).

\bibitem{30}
V.I. Yukalov, {\it Statistical Green's Functions} (Queen's University,
Kingston, 1998).

\bibitem{31}
N.N. Bogolubov, J. Phys. (Moscow) {\bf 11}, 23 (1947).

\bibitem{32}
N.N. Bogolubov, Moscow Univ. Phys. Bull. {\bf 7}, 43 (1947).

\bibitem{33}
N.N. Bogolubov, {\it Lectures on Quantum Statistics} (Gordon and Breach, New
York, 1967), Vol. 1.

\bibitem{34}
V.I. Yukalov, Phys. Rev. E {\bf 72}, 066119 (2005).

\bibitem{35}
V.I. Yukalov, Laser Phys. {\bf 16}, 511 (2006).

\bibitem{36}
V.I. Yukalov and H. Kleinert, Phys. Rev. A  {\bf 73}, 063612 (2006).

\bibitem{37}
V.I. Yukalov, Laser Phys. Lett. {\bf 3}, 406 (2006)

\bibitem{38}
M.D. Girardeau and R. Arnowitt, Phys. Rev. {\bf 113}, 755 (1959).

\bibitem{39}
M.D. Girardeau,  Phys. Rev. {\bf 115}, 1090 (1959).

\bibitem{40}
F. Takano, Phys. Rev. {\bf 123}, 699 91961).

\bibitem{41}
M.D. Girardeau, in {\it Relativity, Particle Physics and Cosmology}, edited
by R.E. Allen (World Scientific, Singapore, 1999), p. 190.

\bibitem{42}
K. Binder and A.P. Young, Rev. Mod. Phys. {\bf 58}, 801 (1986).

\bibitem{43}
V.I. Yukalov and E.P. Yukalova, Phys. Part. Nucl. {\bf 31}, 561 (2000).

\bibitem{44}
V.I. Yukalov and E.P. Yukalova, Phys. Part. Nucl. {\bf 35}, 348 (2004).

\bibitem{45}
V.I. Yukalov, Phys. Rev. B {\bf 71}, 184432 (2005).

\bibitem{46}
N.M. Hugenholtz and D. Pines, Phys. Rev. {\bf 116}, 489 (1959).

\bibitem{47}
J. Gavoret and P. Nozi\`eres, Ann. Phys. (N.Y.) {\bf 28}, 349 (1964).

\bibitem{48}
N.N. Bogolubov, {\it Lectures on Quantum Statistics} (Gordon and Breach, New
York, 1970), Vol. 2.

\bibitem{49}
C.W. Gardiner, {\it Handbook of Stochastic Methods} (Springer, Berlin, 1985).

\bibitem{50}
G.A. Baker and P. Graves-Moris, {\it Pad\'e Approximants} (Cambridge
University, Cambridge, 1996).

\bibitem{51}
V.I. Yukalov, Phys. Rev. A {\bf 42}, 3324 (1990).

\bibitem{52}
V.I. Yukalov, J. Math. Phys. {\bf 32}, 1235 (1991).

\bibitem{53}
V.I. Yukalov, J. Math. Phys. {\bf 33}, 3994 (1992).

\bibitem{54}
V.I. Yukalov, S. Gluzman, and D. Sornette, Physica A {\bf 328}, 409 (2003).

\bibitem{55}
S. Gluzman, V.I. Yukalov, and D. Sornette, Phys. Rev. E {\bf 67}, 026109
(2003).

\bibitem{56}
V.I. Yukalov and S. Gluzman, Int. J. Mod. Phys. B {\bf 18}, 3027 (2004).

\bibitem{57}
E.C. Titchmarsh, {\it Introduction to the Theory of Fourier Integrals} (Oxford
University, Oxford, 1937).

\bibitem{58}
E.C. Titchmarsh, {\it The Theory of Functions} (Oxford University, Oxford,
1939).

\bibitem{59}
V.I. Yukalov and S. Gluzman, Phys. Rev. E {\bf 55}, 6552 (1997).

\bibitem{60}
V.I. Yukalov and S. Gluzman, Phys. Rev. E {\bf 58}, 1359 (1998).

\end{thebibliography}
\end{document}